# Defects induced ferromagnetism in plasma-enabled graphene nanopetals

Z. J. Yue,[1,2] D. H. Seo,[2,3] K. Ostrikov[1,2,3] and X. L. Wang[1*]

[1] Institute for Superconducting and Electronic Materials (ISEM), Faculty of Engineering, University of Wollongong, NSW 2522, Australia

[2] Plasma Nanoscience Centre Australia (PNCA), CSIRO Materials Science and Engineering, P. O. Box 218, Lindfield, NSW 2070, Australia

[3] Plasma Nanoscience, School of Physics, The University of Sydney, Sydney, NSW 2006, Australia.

*Corresponding author: xiaolin@uow.edu.au

Ferromagnetism in graphene is fascinating, but it is still a big challenge for practical applications due to the weak magnetization. In order to enhance the magnetization, here, we design plasma-enabled graphene nanopetals with ultra-long defective edges of up to $10^5$ m/g, ultra-dense lattice vacancies and hydrogen chemisorptions. The designed graphene nanopetals display robust ferromagnetism with large saturation magnetization of up to 2 emu/g at 5 K and 1.2 emu/g at room temperatures. This work identifies the plasma-enabled graphene nanopetals as a promising candidate for graphene-based magnetic devices.

Magnetic ordering in carbon-based materials has attracted widespread attention in scientific communities due to only light elements rather than transition metals like Fe, Co and Ni involved.[1,2-3] Carbon-based magnetic materials possess lots of attractive properties, such as low density, biocompatibility, plasticity and so on, which are practical significance for next generation magnetic devices.[3-4] Graphene, a 2D single sheet of carbon, shows extraordinarily high electron mobility, thermal conductivity and mechanical strength, and holds great promising for many applications such as nanoelectronics, spintronics and optoelectronics.[5] Magnetic ordering in such graphene may have great potential to be used in the design of future magnetic nanoelectronic and spintronic devices. Ferromagnetism has been predicted and observed in graphite and graphene, which is attributed to localized unpaired spins induced by defects.[4,6,7-8] The induced magnetic moments interact ferromagnetically if defects locate at different hexagonal sublattices of graphene.[9]

Many approaches have been used to produce ferromagnetism in graphene, including introduction of hydrogen (H) chemisorptions defects, vacancy defects and edge defects with H plasma process or high-energy ions irradiation.[8,10] To date, however, the observed maximum magnetization in either polycrystalline or single crystal graphene synthesized by CVD or mechanical cleavage is weak.[3,7] The lack of large magnetization hinders the applications of graphene in practical magnetic devices. On the other hand, assuming each carbon atom has a ferromagnetic moment of 1 μB, graphene would have magnetization $\sigma_s$ = 465 emu/g, which means only a tiny fraction of the carbon atoms participate in the magnetism.[2] In order to achieve large magnetization and realize practical application, it is essential to employ new techniques to enhance intrinsic magnetization of graphene.

Here, to achieve robust ferromagnetism and large saturation magnetization, we design graphene nanopetals (vertically aligned petal-like graphene) with ultra-long defective edges, ultra-dense vacancies and H chemisorptions on edges, vacancies and carbon (C) atoms. Indeed, plasma-enabled graphene nanopetals display robust ferromagnetism with enhanced saturation magnetization of up to 2 emu/g at low temperatures and 1.2 emu/g at room temperatures. The observed ferromagnetism is intrinsic magnetic behavior of graphene and can be attributed to the ulta-long defective edges, plasma induced vacancy defects and H adatoms in growth process of graphene nanopetals. Our results indicate plasma aided synthesis is a new way to enhance the ferromagnetism in graphene for practical applications.

Plasma aided growth has been found quite suitable for the controlled synthesis of nanostructures.[11] Vertical graphene networks are grown directly with self-organized patterns by the PECVD without any catalyst and substrate heating.[12] The growth of vertical graphene nanosheets was carried out in an inductively coupled plasma-enhanced chemical vapour deposition (ICP-CVD, 13.56 MHz, 1.0 kW) reactor. Firstly, Ar gas was fed into the chamber where plasma was generated at 3.0 Pa with RF power 800 W. And -50 V of substrate bias was used to enhance the plasma interaction with Si substrate. After 3 minutes of plasma treatment, a gas mixture of 30% $CH_4$, 20% $H_2$ and 50% Ar was fed into the chamber and the rf power increased up to 800 W for the deposition. Here, H atoms can be adsorbed on graphene surface when H plasma is exposed to the sample at high temperatures. Hence, plasma serves as a main H source and produces H absorption defects in graphene [13-14]. The deposition time was kept for 8 minutes.

Figures 1a shows the scanning electron microscope (SEM) images of plasma enabled graphene nanopetals on Si, in which graphene nonosheets are all nearly perpendicular to their substrates, with height ranging from 300 to 500 nm. Based on the scale of SEM images, the estimated edge length is up to $10^5$ m/g. To ensure the magnetic signal originating from the graphene nanosheets, we have analyzed the vertical graphene with x-ray photoelectron spectroscopy (XPS) and Energy-dispersive X-ray spectroscopy (EDX). As shown in Fig. 1b,c only carbon elements have been detected, which proves that our graphene nanopetals have high purity.

The magnetization analysis of the vertical graphene films has been performed with vibrating sample magnetometry (VSM) in Physical Property Measurement System (PPMS). To exclude the influences of $SO_2$/Si, the plasma enabled graphene nanopetals were cleaved from the substrate before VSM measurements. Firstly, the sample holder and plastic tapes which were used to carry samples were measured and display absolute diamagnetism, as shown in Fig. 2a. Fig. 2b displays the magnetic hysteresis loops of the plasma-enabled graphene powders (milled from graphene nanopetals) measured at temperatures 5, 50 and 200 K, which show weak ferromagnetism at 5 K and paramagnetism at higher temperatures. The saturation magnetization $M_s$ is about 0.6 emu/g at 200 K.

Figure 3 exhibits the magnetization hysteresis loops for the plasma-enabled graphene nanopetals at 5, 10, 50, 100, 200 and 300 K. The saturation magnetic moment reaches up to 2 emu/g at 5 K and 1.2 emu/g at 300 K. The ferromagnetism can be seen clearly in the enlarged hysteresis loops in Fig. 3b. Fig. 4a shows the magnetic behavior of magnetic graphite as a function of temperatures [zero field cooled (ZFC) and field cooled (FC) for H=1 kOe]. A clear magnetic transition was observed at $T_c$ = 365 K. The extracted temperature dependence of saturation magnetization and coercivity are plotted in Fig. 4b and c. Compared with the magnetization in plasma-enabled graphene powder, the saturation is giant enhanced. This shows that morphology and structures of graphene nanopetals play a significant role in the generation of robust ferromagnetism and giant saturation magnetization.

As it is well known, H adtoms can generate spontaneous magnetism in graphene.[15] In general, H atoms adsorb on graphite surface via the formation of stable H clusters consisting of two to four H atoms.[16] H adtoms break the double C=C bond and format C-H bond, and release unpaired electrons. The bonding of H and C atoms results in a removal of the π orbital from the low energy sector. Additionally, the H chemisorption also leads to a transition from $sp^2$ to $sp^3$ - $sp^2$ hybridization and $sp^3$-type defects introduce local sublattice imbalances and unpaired spin electrons.[13, 17] There are two carbon atoms per unit cell located at two inequivalent sublattices, A and B. The H chemisorption defects give rise to the strong Stoner ferromagnetism with a magnetic moment of 1 $\mu_B$ per defect when the defects are located at the same sublattice.[9,18] The stability of the magnetic configurations depends on the distance between H adatoms and the strength of exchange couplings between the defect-induced magnetic moments.[18] As shown in Fig. 5, in plasma-enabled graphene nanopetals, the H adatoms are not only adsorbed on the graphene surfaces, but also at interlayers in the plasma-growth process.[19] Partial H adatoms induce the formation of unpaired electrons which lead to the observed magnetization.[20] Theoretical calculations suggest that H maintains the magnetic moment of the defects and give rise to a macroscopic magnetic signal.[21]

The long-range coupling of local moments is expected to take place through spin alternation due to the presence of half-filled π-orbitals in graphene.[22] DFT calculations show that periodic H adsorption results in electronic band structure with the highest occupied band filled with electrons of same spin in a ferromagnetic configuration.[23] As calculated by density functional theory, the magnetic moments interact, ferromagnetism or antiferromagnetism, depend on the relative adsorption graphene sublattice.[24] Strong long-range coupling between local magnetic moments at same sublattice can maintain room-temperature ferromagnetic ordering against thermal fluctuations. The stability of the magnetic configurations depends on the distance between H atoms and the strength of exchange couplings between the defect-induced magnetic moments. The H induced magnetic moments can be aligned by magnetic fields and generate ferromagnetic state that has maximized exchange energy. H saturation can stabilize the vacancy structure and induce magnetic coupling between the defects and the ferromagnetic ordering is accompanied with a semiconducting property.[25] Ferromagnetic behavior in plasma-enabled graphene can be attributed to topological defects and strong Coulomb interaction between electrons [26].

The zigzag edges in graphene nanoribbons can be magnetic due to introducing spin polarized edge states. The net spin moment in zigzag-edged nanosheets results from topological frustration of the π-bonds.[27] In our case, the plasma-enabled graphene nanopetals have one side with open defective edge and another side attached on substrates. The zigzag and armchair edges coexist in plasma-enabled nanopetals. The absorption of H at the edges of graphene nanopetals leads to the formation of a spin-polarized band at the Fermi level. The π electrons on hydrogenated zigzag edges may create a ferromagnetic spin structure on the edge due to edge localized states. H adatoms at zigzag edges can passivate the σ dangling bonds and leave all the π orbitals unsaturated and carrying the magnetic moments. Adsorption of atomic H in graphene leads to a magnetic moment of 1 μB localized on the orbitals surrounding each H atom. In the vertical graphene nanosheets, there is ultra-long edges of up to $10^5$ m/g, which results in giant saturation magnetization. The magnitude of the ferromagnetism strongly depend on the position of the H atoms relative to the edges.[24]

Charged plasma aided growth of graphene nanopetals give rise to the formation of large amount of atomic vacancies in the graphene nanosheets. These lattice vacancies also generate localized electronic states and magnetic moments due to the hybridization of $p_z$ orbitals in the π-band. Calculated magnetic moments are 1.12~1.53 μB per vacancy defect depending on the defect concentration.[9] And vacancy defects in the same sublattice induce ferromagnetic ordering based on theoretical calculation. But the naked vacancy defects generate much weaker ferromagnetic order than H-defect ones. On the other hand, H adatoms can easily adsorb on vacancy dangling bonds and format vacancy-H complexes that can provide larger magnetic moment, which might plays a dominant role in plasma-enabled graphene nanosheets.[21] Density functional theory (DFT) calculations show that H saturation stabilizes vacancy-induced ferromagnetic ordering in graphene, which has to be accompanied with a semiconducting property.[25]

In conclusion, the plasma-enabled graphene nanopetals exhibit robust ferromagnetic ordering with giant enhanced saturation magnetization. The observed ferromagnetism is much higher than the obtained magnetization in the case of flat few-layer graphene. Plasma-enabled graphene nanopetals can be particularly useful to explore magnetic ordering in graphene through introducing controllable edge, lattice vacancy and H defects. By manipulating the conditions of plasma, it is possible to tune flexibly the magnetism of graphene. The plasma-enabled nanopetals are very promising candidates for the graphene based magnetic nanodevices.

**Acknowledges**

This work is partially supported the Australian Research Council under Discovery Project DP1094073 and Future Fellowship Project FT100100303.

**Reference**

1. R. C. Haddon, Nature **378** (6554), 249 (1995); Tatiana L. Makarova, Bertil Sundqvist, Roland Hohne, Pablo Esquinazi, Yakov Kopelevich, Peter Scharff, Valerii A. Davydov, Ludmila S. Kashevarova, and Aleksandra V. Rakhmanina, Nature **413** (6857), 716 (2001); L. Krusin-Elbaum, D. M. Newns, H. Zeng, V. Derycke, J. Z. Sun, and R. Sandstrom, Nature **431** (7009), 672 (2004).
2. J. M. D. Coey, M. Venkatesan, C. B. Fitzgerald, A. P. Douvalis, and I. S. Sanders, Nature **420** (6912), 156 (2002).
3. Oleg V Yazyev, Reports on Progress in Physics **73** (5), 056501 (2010).
4. Donghan Seo, Zengji Yue, Xiaolin Wang, Igor Levchenko, Shailesh Kumar, Shixue Dou, and Kostya Ostrikov, Chemical Communications **49** (99), 11635 (2013).
5. A. H. Castro Neto, F. Guinea, N. M. R. Peres, K. S. Novoselov, and A. K. Geim, Reviews of Modern Physics **81** (1), 109 (2009); M. O. Goerbig, Reviews of Modern Physics **83** (4), 1193 (2011).
6. J. Cervenka, M. I. Katsnelson, and C. F. J. Flipse, Nat Phys **5** (11), 840 (2009); Chandra Sekhar Rout, Anurag Kumar, Nitesh Kumar, A. Sundaresan, and Timothy S. Fisher, Nanoscale **3** (3), 900 (2011); P. Esquinazi, D. Spemann, R. Höhne, A. Setzer, K. H. Han, and T. Butz, Physical Review Letters **91** (22), 227201 (2003); Physical Review B **83** (15), 155445 (2011).
7. Yan Wang, Yi Huang, You Song, Xiaoyan Zhang, Yanfeng Ma, Jiajie Liang, and Yongsheng Chen, Nano Letters **9** (1), 220 (2008).
8. Huihao Xia, Weifeng Li, You Song, Xinmei Yang, Xiangdong Liu, Mingwen Zhao, Yueyuan Xia, Chen Song, Tian-Wei Wang, Dezhang Zhu, Jinlong Gong, and Zhiyuan Zhu, Advanced Materials **20** (24), 4679 (2008).
9. Oleg V. Yazyev and Lothar Helm, Physical Review B **75** (12), 125408 (2007).
10. A. V. Krasheninnikov and F. Banhart, Nat Mater **6** (10), 723 (2007); Andreas Ney, Pagona Papakonstantinou, Ajay Kumar, Nai-Gui Shang, and Nianhua Peng, Applied Physics Letters **99** (10), 102504 (2011); David Soriano, Nicolas Leconte, Pablo Ordejón, Jean-Christophe Charlier, Juan-Jose Palacios, and Stephan Roche, Physical Review Letters **107** (1), 016602 (2011).
11. K. Ostrikov, Reviews of Modern Physics **77** (2), 489 (2005); Kostya Ostrikov and Hamid Mehdipour, Journal of the American Chemical Society **134** (9), 4303 (2012); Shailesh Kumar, Hamid Mehdipour, and Kostya Ostrikov, Advanced Materials, n/a (2012).

[12] D. H. Seo, S. Kumar, and K. Ostrikov, Carbon **49** (13), 4331 (2011); Dong Han Seo, Shailesh Kumar, and Kostya Ostrikov, Journal of Materials Chemistry **21** (41), 16339 (2011); Shailesh Kumar and Kostya Ostrikov, Nanoscale **3** (10), 4296 (2011).

[13] P. Ruffieux, O. Gröning, M. Bielmann, P. Mauron, L. Schlapbach, and P. Gröning, Physical Review B **66** (24), 245416 (2002).

[14] P. Ruffieux, O. Gröning, P. Schwaller, L. Schlapbach, and P. Gröning, Physical Review Letters **84** (21), 4910 (2000); D. C. Elias, R. R. Nair, T. M. G. Mohiuddin, S. V. Morozov, P. Blake, M. P. Halsall, A. C. Ferrari, D. W. Boukhvalov, M. I. Katsnelson, A. K. Geim, and K. S. Novoselov, Science **323** (5914), 610 (2009).

[15] Kathleen M. McCreary, Adrian G. Swartz, Wei Han, Jaroslav Fabian, and Roland K. Kawakami, Physical Review Letters **109** (18), 186604 (2012).

[16] A. Allouche, Y. Ferro, T. Angot, C. Thomas, and J.-M. Layet, The Journal of Chemical Physics **123** (12), 124701 (2005).

[17] J. Zhou, Q. Wang, Q. Sun, X. S. Chen, Y. Kawazoe, and P. Jena, Nano Letters **9** (11), 3867 (2009); Oleg V. Yazyev, Physical Review Letters **101** (3), 037203 (2008); Elliott H. Lieb, Physical Review Letters **62** (10), 1201 (1989).

[18] Ahmad Ranjbar, Mohammad Saeed Bahramy, Mohammad Khazaei, Hiroshi Mizuseki, and Yoshiyuki Kawazoe, Physical Review B **82** (16), 165446 (2010).

[19] Young-Woo Son, Marvin L. Cohen, and Steven G. Louie, Nature **444** (7117), 347 (2006).

[20] Lanfei Xie, Xiao Wang, Jiong Lu, Zhenhua Ni, Zhiqiang Luo, Hongying Mao, Rui Wang, Yingying Wang, Han Huang, Dongchen Qi, Rong Liu, Ting Yu, Zexiang Shen, Tom Wu, Haiyang Peng, Barbaros Ozyilmaz, Kianping Loh, Andrew T. S. Wee, Ariando, and Wei Chen, Applied Physics Letters **98** (19), 193113 (2011).

[21] P. O. Lehtinen, A. S. Foster, Yuchen Ma, A. V. Krasheninnikov, and R. M. Nieminen, Physical Review Letters **93** (18), 187202 (2004).

[22] Jeongmin Hong, Sandip Niyogi, Elena Bekyarova, Mikhail E. Itkis, Palanisamy Ramesh, Nissim Amos, Dmitri Litvinov, Claire Berger, Walt A. de Heer, Sakhrat Khizroev, and Robert C. Haddon, Small **7** (9), 1175 (2011).

[23] Yves Ferro, D. Teillet-Billy, N. Rougeau, V. Sidis, S. Morisset, and Alain Allouche, Physical Review B **78** (8), 085417 (2008).

[24] D. Soriano, F. Muñoz-Rojas, J. Fernández-Rossier, and J. J. Palacios, Physical Review B **81** (16), 165409 (2010).

[25] Weifeng Li, Mingwen Zhao, Xian Zhao, Yueyuan Xia, and Yuguang Mu, Physical Chemistry Chemical Physics **12** (41), 13699 (2010).

[26] P. Esquinazi, A. Setzer, R. Höhne, C. Semmelhack, Y. Kopelevich, D. Spemann, T. Butz, B. Kohlstrunk, and M. Lösche, Physical Review B **66** (2), 024429 (2002).

[27] Wei L. Wang, Sheng Meng, and Efthimios Kaxiras, Nano Letters **8** (1), 241 (2007).

**Figures and captions**

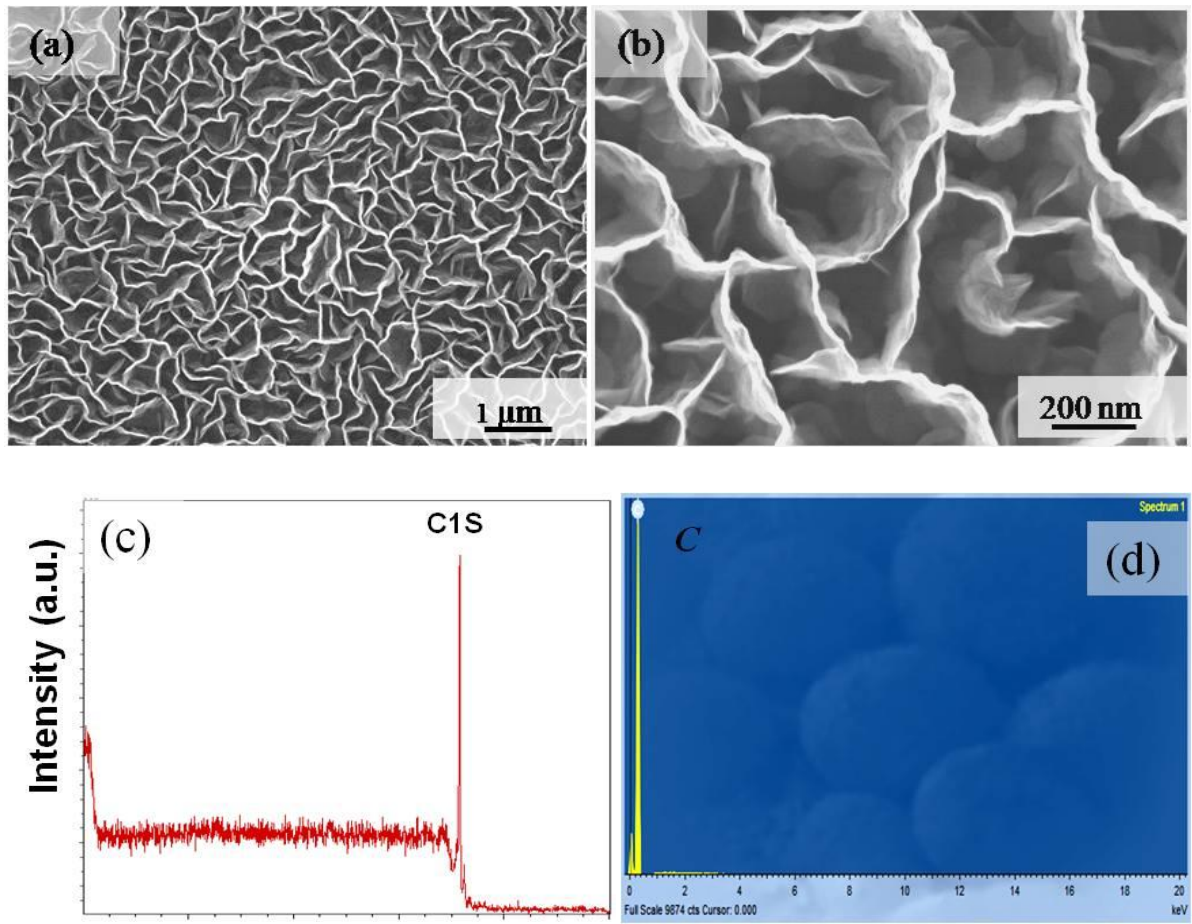

**Figure 1** (a, b) SEM images, (c) XPS spectra and (d) EDX spectra for the plasma-enabled graphene nanopetals.

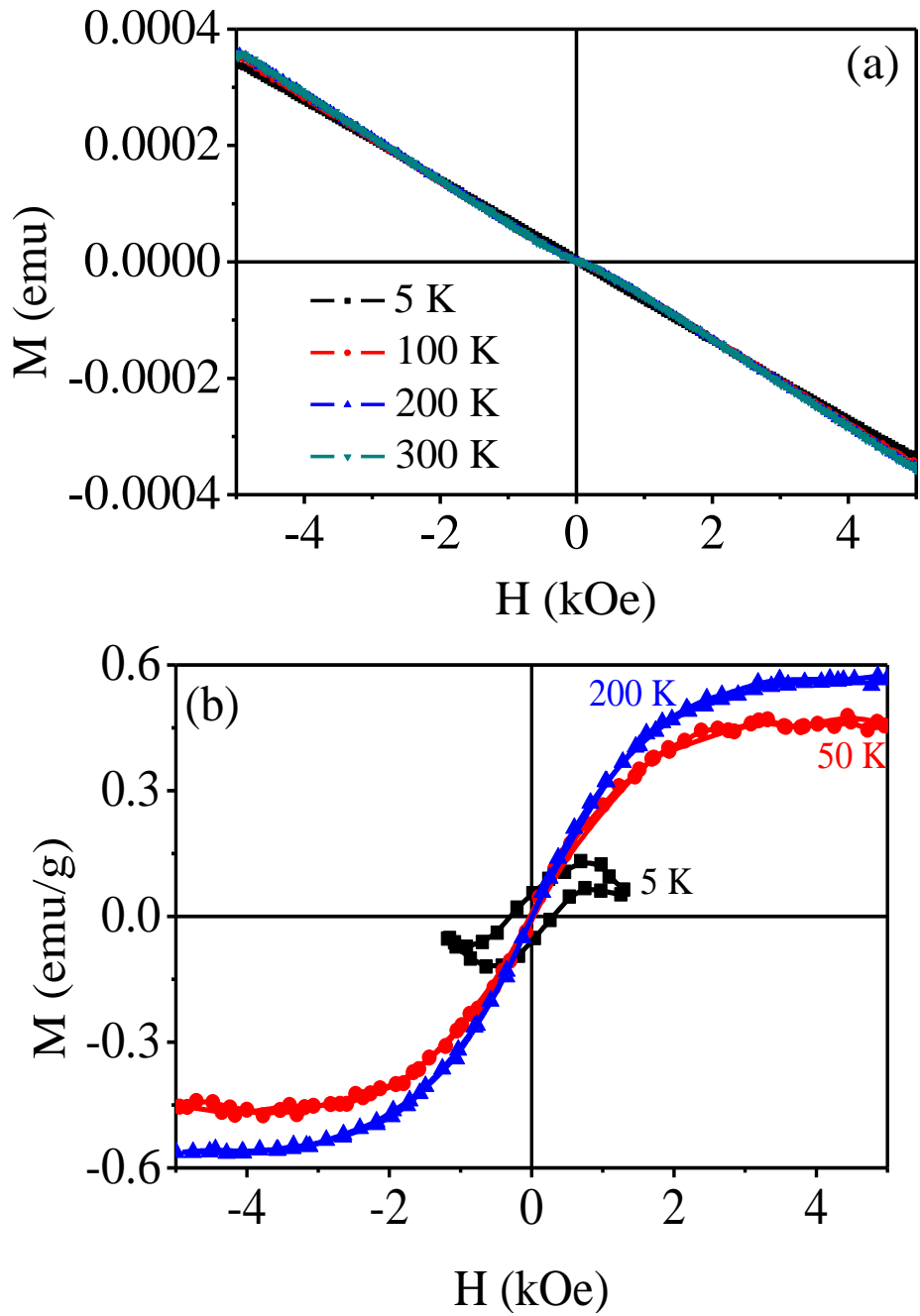

**Figure 2** Magnetization hysteresis loops for (a) sample holder, and (b) plasma-enabled graphene powders.

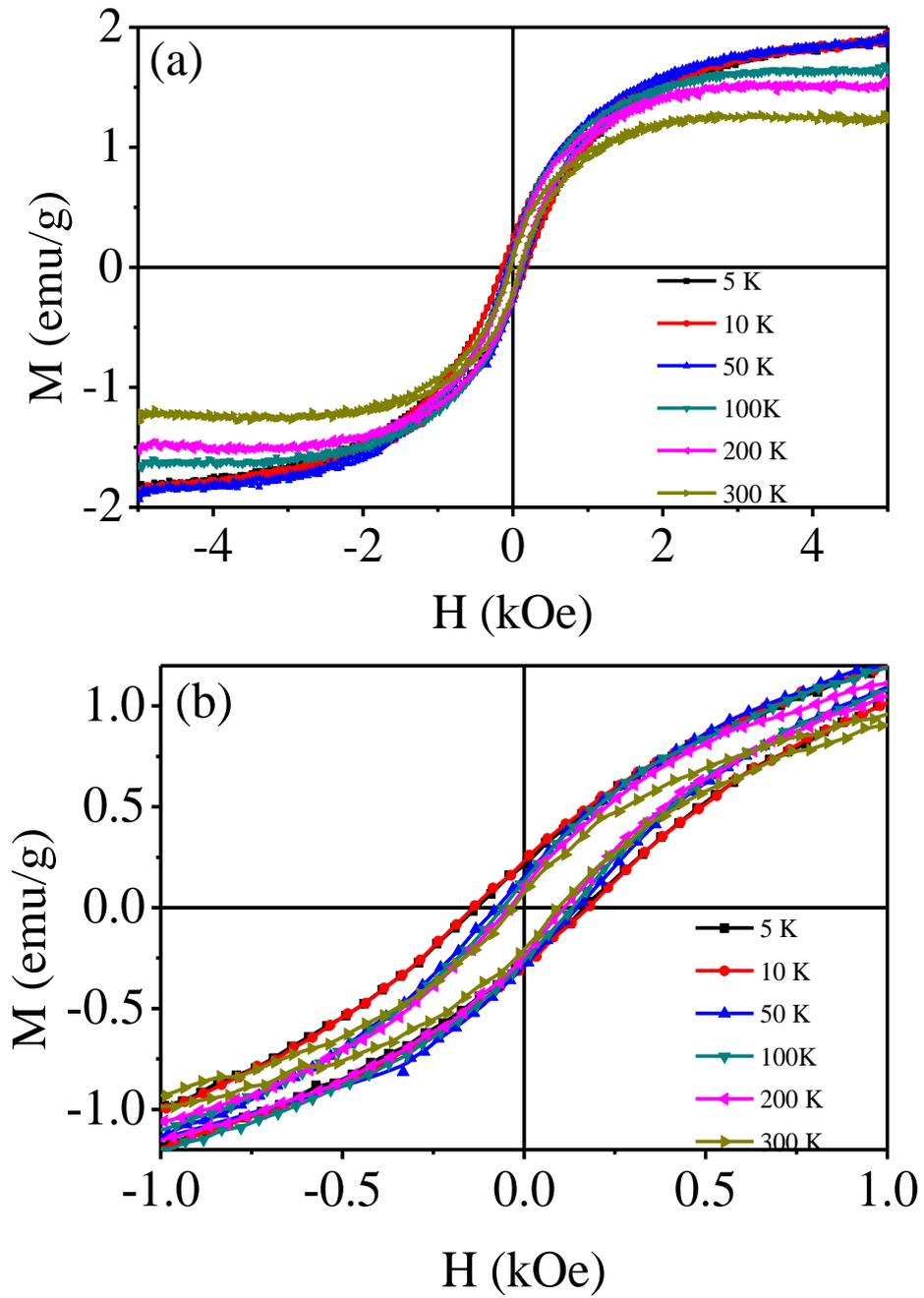

**Figure 3** Magnetization hysteresis loops for plasma-enabled graphene nanopetals at different temperatures.

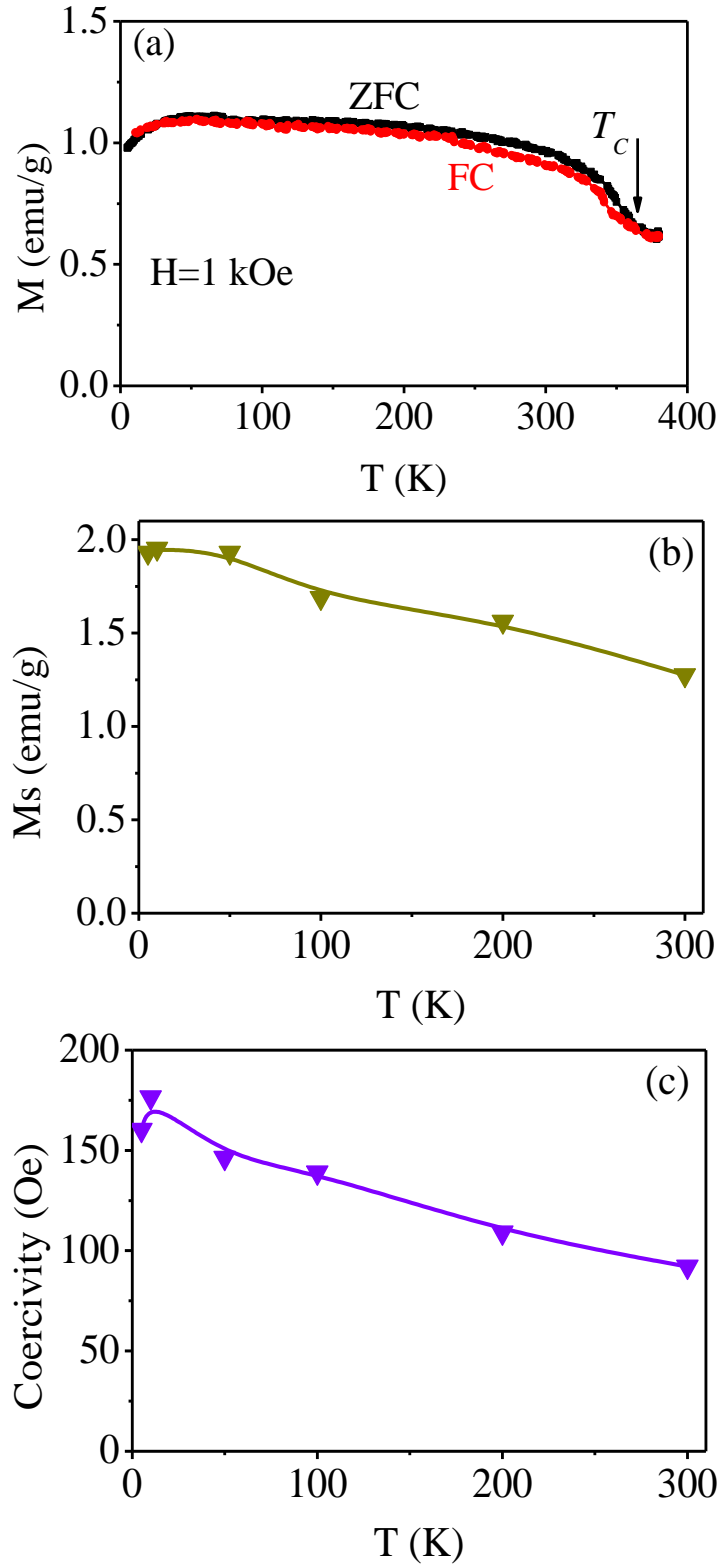

**Figure 4** (a) ZFC and FC magnetization as a function of temperatures, ranging from 5 K to 380 K. The arrow shows the magnetic transition temperatures, $T_c$ = 365 K. (b) Saturation magnetization and (c) Coercivity as a function of temperatures.

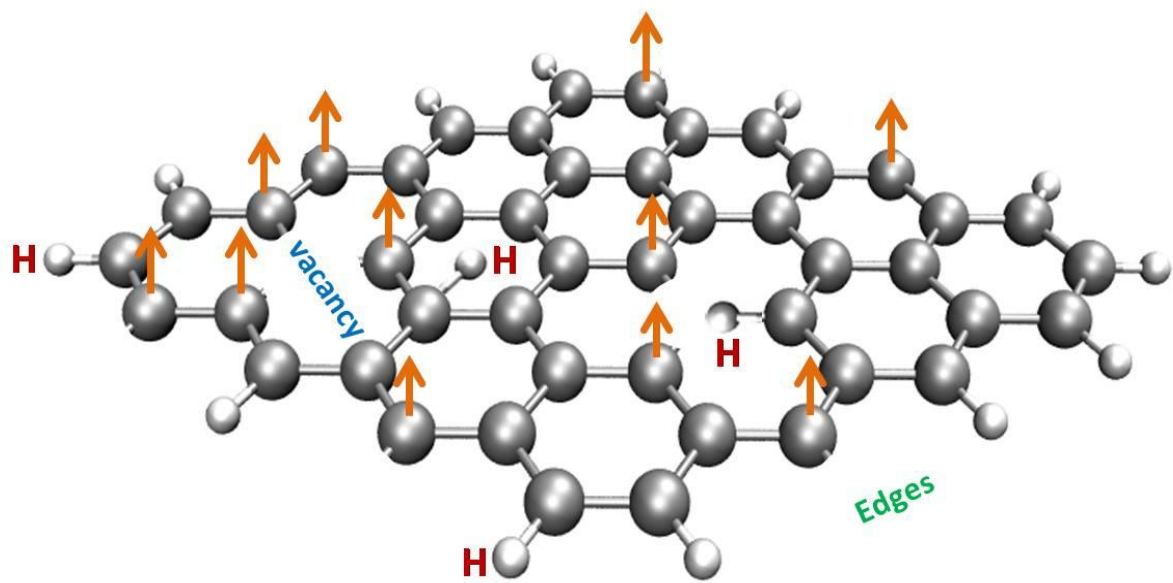

**Figure 5** Atomic structures of the graphene nanopetals with hydrogen chemisorption defects, edge defects and lattice vacancy defects.